# Topology optimization of broadband hyperbolic elastic metamaterials with super-resolution imaging

Hao-Wen Dong[a,b], Sheng-Dong Zhao[a,b], Yue-Sheng Wang[a,*], Chuanzeng Zhang[b,*]

[a]*Institute of Engineering Mechanics, Beijing Jiaotong University, Beijing 100044, China*
[b]*Department of Civil Engineering, University of Siegen, D-57068 Siegen, Germany*

**Abstract**

Hyperbolic metamaterials are strongly anisotropic artificial composite materials at a subwavelength scale and can greatly widen the engineering feasibilities for manipulation of wave propagation. However, limited by the empirical structure topologies, the previously reported hyperbolic elastic metamaterials (HEMMs) suffer from the limitations of relatively narrow frequency width, inflexible adjusting operating subwavelength scale and being difficult to further ameliorate imaging resolution. Here, we develop an inverse-design approach for HEMMs by topology optimization based on the effective medium theory. We successfully design two-dimensional broadband HEMMs supporting multipolar resonances, and theoretically demonstrate their deep-subwavelength imagings for longitudinal waves. Under different prescribed subwavelength scales, the optimized HEMMs exhibit broadband negative effective mass densities. Moreover, benefiting from the extreme enhancement of evanescent waves, an optimized HEMM at the ultra-low frequency can yield a super-high imaging resolution ($\sim\lambda/64$), representing the record in the field of elastic metamaterials. The proposed computational approach can be easily extended to design hyperbolic metamaterials for other wave counterparts. The present research may provide a novel design methodology for exploring the HEMMs based on unrevealed resonances and serve as a useful guide for the ultrasonography and general biomedical applications.

## 1. Introduction

Metamaterials, artificial composite subwavelength materials or structures, provide many encouraging opportunities to modulate and control wave propagation with the extraordinary physical properties. In recent years, due to the strongly anisotropic dispersion, the hyperbolic metamaterials (HMMs) with hyperbolic dispersions [1] as one of the most important types of metamaterials have attracted wide attention and showed many promising applications, including negative refraction [2-4], enhanced superlensing effects [5-12], backward waves [7], strong thermal emission [13-16], sensing [17], Purcell factor enhancement [2, 18], improved absorption [19-21], heightened conductivity [22] and intensified spontaneous emission [2, 23-26]. In particular, compared with the double-negative metamaterials, the HMMs just need to satisfy the criterion about constraining the motion of particles in one or two principle directions [2]. The concept of HMMs has been applied to engineering materials for better controlling the electromagnetic waves [2, 5, 6, 12-27] and acoustic waves [3, 4, 7, 9]. Unlike the electromagnetic [5] and acoustic [29] counterparts, elastic metamaterials (EMMs) [30-34] involve more constitutive parameters, supporting two modes of longitudinal and transverse waves. They can offer more possibilities to explore physical phenomena [30-34] beyond natural materials. Thereby, combing the characteristics of HMMs and EMMs, it is more challenging to construct hyperbolic elastic materials (HEMMs) [8, 10, 11, 28] with a set of desired properties.

* Corresponding author. Tel.: +86 10 51688417; fax: +86 10 51682094. *E-mail address:* yswang@bjtu.edu.cn (Y. S. Wang).
* Corresponding author. Tel.: +49 271 7402173; fax: +49 271 7404074. *E-mail address:* c.zhang@uni-siegen.de (Ch. Zhang).

Over the past two years, several research groups have focused on the HEMMs and proposed different microstructure topologies [8, 10, 11, 28]. Their superlenging capacities have been numerically [8, 10, 11] and experimentally demonstrated [8, 11]. Two different mechanisms [8, 10, 11, 28] were shown to give rise to the hyperbolic dispersions. A coiling-up metamaterial possessing different deformations along two principle directions was reported to bring about the superior imaging resolution which breaks the diffraction limit [8]. Then, an elastic hyperlens was designed based on the microstructure with anisotropic mass densities [10, 11]. Actually, for elastic media, it is much easier to acquire a low-frequency bandgap through anisotropic mass densities than through distinct deformations mechanisms. However, the systematical design of anisotropic mass densities for the hyperbolic dispersion is still missing. Moreover, the reported HEMMs show the following limitations: (1) The operating frequency range is relatively narrow. We have to make further improvements in widening frequency and wave-vector ranges for hyperbolic eigenfrequency curves (EFCs). (2) The HEMMs at ultra-low frequencies have not been implemented. (3) No simple and controllable method can design HEMMs at different subwavelength scales. (4) The imaging resolution of the elastic hyperlens has yet to be improved. We argue that the super-resolution (i.e., much smaller than the diffraction limit) ability of the HEMM brings great challenges to metamaterial engineering. However, the manual and intuitive designs are unable to overcome the above limitations and make the HEMMs more practical. Therefore, a systematic methodology is necessary for seeking the high-performance microstructure topologies exhibiting hyperbolic dispersion in modulating the elastic subwavelength waves.

In this paper, based on the topology optimization [12, 28, 34, 37, 38] and effective medium theory [30, 32], we develop a simple inverse method to realize two-dimensional (2D) broadband HEMMs with negative effective mass densities along one principle direction. We show that the analogical geometry features of the optimized HEMMs and reveal their common multipole resonance mechanism and controlled vibration along the propagating direction. All optimized HEMMs are fully proved to support the subwavelength imaging. In particular, we demonstrate that a single-phase metamaterial with suitable constraints can exhibit the hyperbolic dispersion at the ultra-low frequency range, implying the comparable capacity of manipulating elastic waves as in the multi-phase local resonance metamaterials. As a result, the longitudinal waves only along the desired direction can propagate within the HEMMs. Furthermore, our optimized HEMMs can persistently and intensely enhance the transmission of evanescent waves over a largest wave vector range. And we obtain a super-high, or almost ultimate, imaging resolution (~$\lambda/64$) which represents the record value in the field of EMMs for longitudinal waves.

This paper is organized as follows: Section 2 describes the methodology of topology optimization for single negative effective mass densities. Section 3 successively presents the optimized HEMMs, the physical mechanisms about hyperbolic dispersion, hyperlensing phenomena, extreme anisotropy verification and origin of the super-resolution imaging. Finally, the conclusions are drawn in Sec. 4.

## 2. Methodology

To obtain the hyperbolic dispersion, we have to construct the microstructure with anisotropic mass densities [10] or elastic stiffnesses [11]. In this paper, we consider a metamaterial with the orthotropic symmetry in a square lattice, as shown in Fig. 1(a). Changing the microstructure topology in Fig. 1(b) will induce the possible resonances, leading to an anisotropic dispersion for a certain energy band. As evidenced in the recent work about HEMMs [10], if the dominated mechanism is the anisotropic mass density, the eigenfrequency curves (EFCs) will be elliptical or hyperbolic, as displayed in Fig. 1(c). Especially, when $\rho_{yy}>\rho_{xx}>0$ or $\rho_{xx}>\rho_{yy}>0$, the curve may take shape 1 or 2. However, if the anisotropy becomes stronger, the hyperbolic shape 3 or 4 is possible to occur for the EMM with a single negative mass density $\rho_{yy}<0<\rho_{xx}$ or $\rho_{xx}<0<\rho_{yy}$, like the electromagnetic and acoustic counterparts [2, 9]. To systematically achieve hyperbolic EFCs in a robust way, we will apply topology optimization in this paper to design the unit-cell microstructure with a single negative mass density by considering the design domain illustrated in Fig. 1(b).

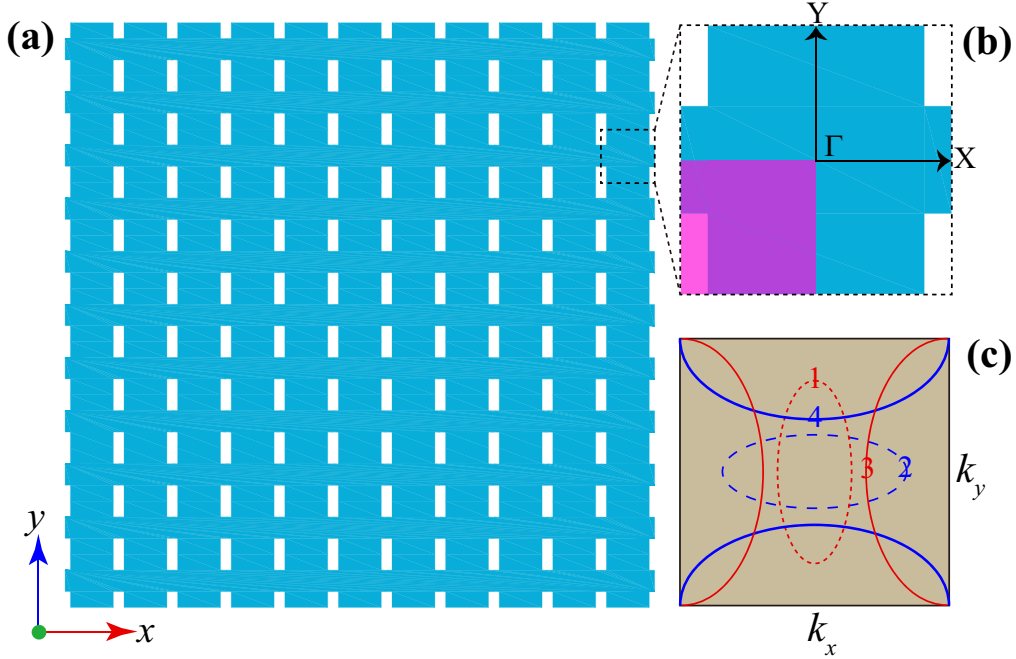

Fig. 1. Schematic illustration of an anisotropic metamaterial. (a) Metamaterial with periodic microstructures. (b) Unit-cell with orthogonal symmetry. (b) Possible EFCs for an anisotropic metamaterial. The principal directions (ΓX and ΓY) of the first Brillouin zone are shown in (b) as well. The unit-cell surrounded by the dashed lines in (b) is taken as the design domain with the shaded area showing the reduced design region in optimization.

It was reported that a negative mass density can be usually realized by dipolar resonances, whereas a negative bulk modulus and a negative shear modulus can originate from monopolar and quadrupolar resonances, respectively [30-32]. Although topology optimization may yield the complex microstructure which goes beyond the existing resonance mechanisms, the whole variation characteristics of the effective parameters are analogical for a metamaterial with either negative mass density or negative elastic modulus. That is, the value reaches the infinity occurs at the resonant frequency, and then gradually decreases with the frequency away from the resonance [30-32]. For simplicity, it is effective to adopt the discrete responses at a certain number of frequencies to define the effective performance. To this end, we select some sampling frequencies distributed uniformly in a target frequency range ($f_{min}$, $f_{max}$). Generally, the essential condition for obtaining a negative mass density is to enlarge the gap between its maximal and minimal positive values with a resonance. Then the negative range can be expanded if the resonant frequency is pushed down. For the same frequency range of interest, the decrease of the resonant frequency will result in the smaller minimal positive value. Consequently, the driving force for the wide enough negative range to a low frequency is to increase the ratio between the positive maximal value to the positive minimal value at all sampling frequencies. For the broadband negative mass density along $x$-direction, we propose the formulation within ($f_{min}$, $f_{max}$) as

$$\text{Maximize}: SN = N - \frac{\min_{\forall m(m \subset (1,2\cdots M))}\left(\rho_{xx}^{+}(m)\right)}{\max_{\forall m(m \subset (1,2\cdots M))}\left(\rho_{xx}^{+}(m)\right)}, \tag{1a}$$

Subject to

$$: \min_{\forall i}(\rho_{yy}(i)) > 0, \tag{1b}$$

$$: \min_{\forall i}(E_{xx}(i)) > 0, \tag{1c}$$

$$: \min_{\forall i}\left(\frac{E_{yy}(i)}{E_{xx}(i)}\right) > 1.0, \tag{1d}$$

$$: \min_{\forall i}\left(\frac{E_{xy}(i)}{E_{yy}(i)}\right) \geq \delta_{E}, \tag{1e}$$

$$: \max_{\forall i}\left(\left(\frac{\sum|F_{x}|}{\sum|F_{y}|}\right)_{i}\right) \leq \delta_{F}, \tag{1f}$$

$$: \frac{\max_{\forall i}(\rho_{yy}(i))}{\min_{\forall i}(\rho_{yy}(i))} \leq \delta_{\rho}, \tag{1g}$$

$$: \min_{\phi}(e) \geq e^{*}, \tag{1h}$$

where $M$ is the number of the sampling frequencies; $SN$ denotes the value of the objective function value; $N$ is the number of the sampling frequency where negative $\rho_{xx}$ is generated; $\rho_{xx}$ ($\rho_{yy}$) and $E_{xx}$ ($E_{yy}$) are the effective mass density and effective stiffness along $x$- ($y$-) directions, respectively; $\rho_{xx}^{+}$ represents the special array composed of the positive values; $m$ ($m \leq M$) is the serial number of the frequency where positive $\rho_{xx}$ remains; $i$ ($i=1, 2\ldots M$) is the serial number of the calculated frequencies; $E_{xy}$ is the coupling stiffness; $F_x$ ($F_y$) is the magnitude of the reaction force along $x$- ($y$-) direction when calculating $\rho_{yy}$; $\Sigma$ means the integration over the upper and lower boundaries of the unit-cell; $e$ is the array composed of the width of every solid connection; and $\delta_E$, $\delta_F$, $\delta_\rho$ and $e^*$ are the self-defined parameters. Constraints (1b)-(1d) are introduced to ensure the emergency of the hyperbolic dispersion. Constraint (1e) is applied to control coupling between the longitudinal waves along two principal directions. Constraint (1f) is used for deleting the strong local rotation. Constraint (1g) is employed to ensure the extreme anisotropy of the effective mass densities. Constraint (1h) is utilized to make the structure manufacturable. Our numerical tests show that $M$=11 can effectively describe the dynamic continuous characters over a frequency spectrum. More sampling frequencies result in the higher computational cost although the optimized results are the same. For all designs with negative $\rho_{xx}$, we take $\delta_F$=0.2, $\delta_\rho$=1.37 and $e^*$=0.001 m based on the numerical tests. The single-objective genetic algorithm (GA) will be adopted [36] to achieve the optimized HEMM for a given frequency range ($f_{\min}$, $f_{\max}$). The kernel of the present design method is that GA generates various microstructure topologies whose effective material parameters are extracted by the effective medium theory to hunt for better objective properties. The details of retrieving effective parameters, objective functions, constraints and GA implementations are presented in the Supplemental Material [36].

## 3. Numerical results and discussions

*3.1. Optimized metal metamaterials*

We consider the design of a square-latticed (lattice constant is $a$=0.03 m) perforated single-phase metal structure made of stainless steel [10, 32, 34] with the mass density $\rho$=7850 kg m$^{-3}$, the Young's modulus $E$=200 GPa and the Poisson's ratio $\upsilon$=0.3. By employing topology optimization, we construct some novel microstructure topologies which are difficult to conceive through conventional intuition. Meanwhile, these distinct HEMMs show the outstanding frequency bandwidths and profoundly reveal some exotic mechanisms for the hyperbolic dispersion. Figure 2 shows the optimized microstructures S1, S2 and S3 under the different wavelength scales of $\lambda_1$=10$a$ ($f_{max}$= 19.5 kHz), $\lambda_2$=20$a$ ($f_{max}$=9.75 kHz) and $\lambda_3$=50$a$ ($f_{max}$=3.904 kHz), respectively. It is noted that the optimized solutions can guarantee all negative $\rho_{xx}$ within ($f_{min}$, $f_{max}$). Since we expect to obtain the wide enough hyperbolic range at low frequencies as much possible, $f_{min}$ for all optimization cases is selected as 0.5 Hz. Besides, we use different $\delta_E$ to felicitously confine the coupling stiffness to ensure the emergence of negative $\rho_{xx}$ in the given search space, especially for the cases with ultra-low frequencies (the effect of $\delta_E$ is discussed in the Supplemental Material [36]). It is seen from Fig. 2 that the three ranges of negative $\rho_{xx}$ are nearly consistent with the relevant frequency ranges in the band structures. The negative ranges with mid-frequencies of 15.3025 kHz, 7.177 kHz and 3.0499 kHz provide the widths of 12.899 kHz, 5.198 kHz and 1.7846 kHz, respectively. It is observed that three microstructures S1, S2 and S3 achieve the noticeable negative ranges not only within but also outside the target range ($f_{min}$, $f_{max}$). In particular, the microstructure S1 produces the simultaneous negative $\rho_{xx}$ and $E_{xx}$ in (19.791 kHz, 21.732 kHz) owing to the multipolar resonances. This means that the optimization objective function in Eq. (1a) successfully drives evolution to generate the wide enough negative range over the whole spectrum instead of restricting to ($f_{min}$, $f_{max}$) where the dispersion relation is hyperbolic.

Noticeably, three fittest microstructures in Fig. 2 exhibit the common geometrical features: (i) multiple solid blocks interconnected by narrow solid connections, (ii) two centered blocks placed in $y$-direction, and (iii) several slender rods locat along $x$- or $y$-direction, i.e., act as either the horizontal or vertical connections. Intuitively, these features will be responsible for the strong anisotropy of dynamic wave responses with respect to two principle directions. Generally speaking, increasing the thickness of the connections gives rise to the increase of the operating frequency range. On the other hand, the anisotropic degree of the effective dynamic behaviors mainly depends on the symmetries and topologies of the multiple blocks.

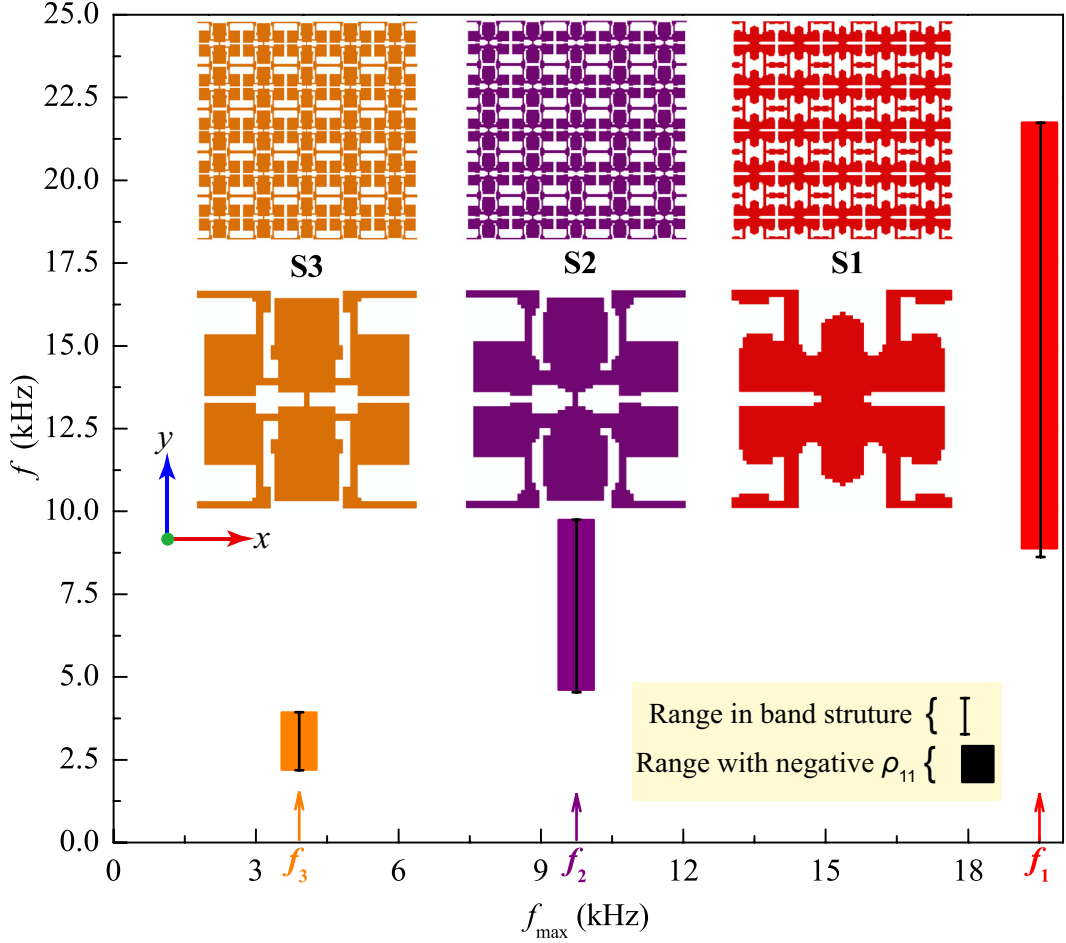

Fig. 2. Optimized HEMMs and representative EFCs. (a) Optimized microstructures for different target frequencies ($f_{max}= f_1, f_2$ and $f_3$). Note that the optimization parameters for microstructures S1, S2 and S3 are selected as ($f_{max}=f_1$=19.5 kHz, $\delta_E$=0.1), ($f_{max}=f_2$=9.75 kHz, $\delta_E$=0.1) and ($f_{max}=f_3$=3.904 kHz, $\delta_E$=0.05) respectively. Their operating wavelengths are $\lambda_1=10a$, $\lambda_2=20a$ and $\lambda_3=50a$, respectively. The solid rectangles represent the lower and upper frequencies of range with negative $\rho_{xx}$. The line bars show the relevant frequency ranges in band structures. Their corresponding 3×3 lattice structures are shown above.

To check the convergence of the calculation, we present in Fig. 3 the evolutionary history of the maximal fitness as a function of the generation number for S3. The metal structure without any void is taken as the initial "seed" structure, see snapshot of generation 0 (G=0). From generation G=0 to G=61, the snapshots show that GA can quickly capture the beneficial geometry at the early evolution, i.e., thin left and right boundaries, to increase the gap between minimal positive $\rho_{xx}$ and maximal positive $\rho_{xx}$. The change from generation G=61 to G=145 means that the solid blocks are useful to further extend the gap. A significant improvement of *SN* takes place from generation G=145 (*SN*=-0.7732) to G=209 (*SN*=0.7663). This implies that the geometry of the narrow boundaries with four big solid blocks and two small solid lumps in the center is effective to generate negative $\rho_{xx}$ at ultra-low frequencies. When the local connections between blocks and boundaries get narrower from generation G=209 (*SN*=0.7663) to G=229 (*SN*=2.6583), the frequency range with negative $\rho_{xx}$ becomes wider. Then, this increasing tendency can be enhanced by removing more solids of connections from generation G=229 (*SN*=2.6583) to G=305 (*SN*=3.631). The negative ranges will increase from generation G=305 (*SN*=3.631) to G=553 (*SN*=4.636) because of the growth of two central lumps. At this time, the evolution in the coarse grid begins to converge. Mapping the topology from a coarse grid into a finer one will lower *SN* at some extent. In the fine grid, GA can commendably make the structural edges clearer and more smooth. From generation G=1004 (*SN*=4.612) to G=2000 (*SN*=4.7623), the structure turns to possess the extreme narrow connections and large enough solid blocks. Overall, evolutionary history in Fig. 3 clearly demonstrates the significance of typical structural features on generation of broadband negative $\rho_{xx}$.

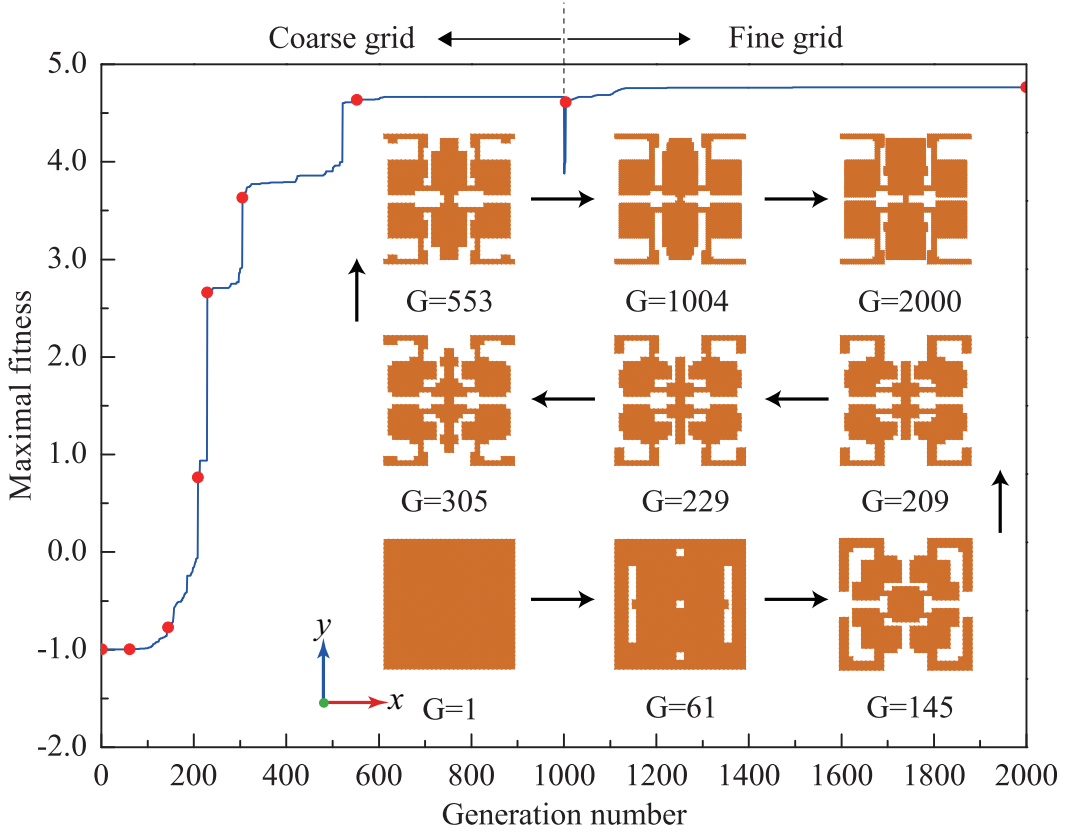

Fig. 3. Evolutionary history for generation of HEMM S3 in Fig. 2. Snapshots of representative topologies during optimization are included. The objective function values (*SN*) of nine microstructures are -0.9988 (G=0), -0.9969 (G=61), -0.7732 (G=145), 0.7663 (G=209), 2.6583 (G=229), 3.631 (G=305), 4.636 (G=553), 4.612 (G=1004) and 4.7623 (G=2000), respectively.

*3.2. Negative properties and mechanism analysis*

To demonstrate the negative properties of the optimized HEMMs, we numerically evaluate the dispersion relation and transmission, and extract the effective parameters for S3, as illustrated in Figs. 4. In Fig. 4(a), we use $q_x = \left|\sum_{\text{unitcell}} u_x\right| / \sqrt{\left(\sum_{\text{unitcell}} u_x\right)^2 + \left(\sum_{\text{unitcell}} u_y\right)^2}$ to characterize the proportion of *x*-polarized vibration for the existing propagating modes. A wide directional bandgap in ΓX direction is opened between two longitudinal wave bands, while a relatively straight longitudinal wave band is maintained within the same range along ΓY direction. The hyperbolic dispersion will be resulted from the different characteristics along orthogonal principle directions of the HEMM microstructure. We present longitudinal wave transmission property based on the optimized HEMM S3 in Fig. 4(b). The results clearly show the total transmission along ΓY direction and total prevention in ΓX direction within the bandgap range. As shown in Figs. 4(c) and 4(d), the HEMM exhibits the apparent anisotropy of the effective mass density and elastic stiffness. When $\rho_{yy}$ keeps nearly a constant value, $\rho_{xx}$ turns to be negative within the range of (2.157 kHz, 3.943 kHz) which matches well with the bandgap in Fig. 4(a). Meanwhile, $E_{xx}$ is always positive in the same range. The HEMM takes the simultaneous positive $\rho_{xx}$ and $E_{xx}$ in the range of (3.94 kHz, 4.385 kHz), which characterizes accurately the modes of the forth band in ΓX direction. Here, we apply the effective longitudinal modulus *P* (see the retrieves of the effective parameters in the Supplemental Material [36]) to characterize the whole effective behaviors concerning the longitudinal wave motion. Unlike $E_{xx}$ and $E_{yy}$, we find no large decrease of $P$. This, in turn, explains the strong anisotropy of the elastic stiffness.

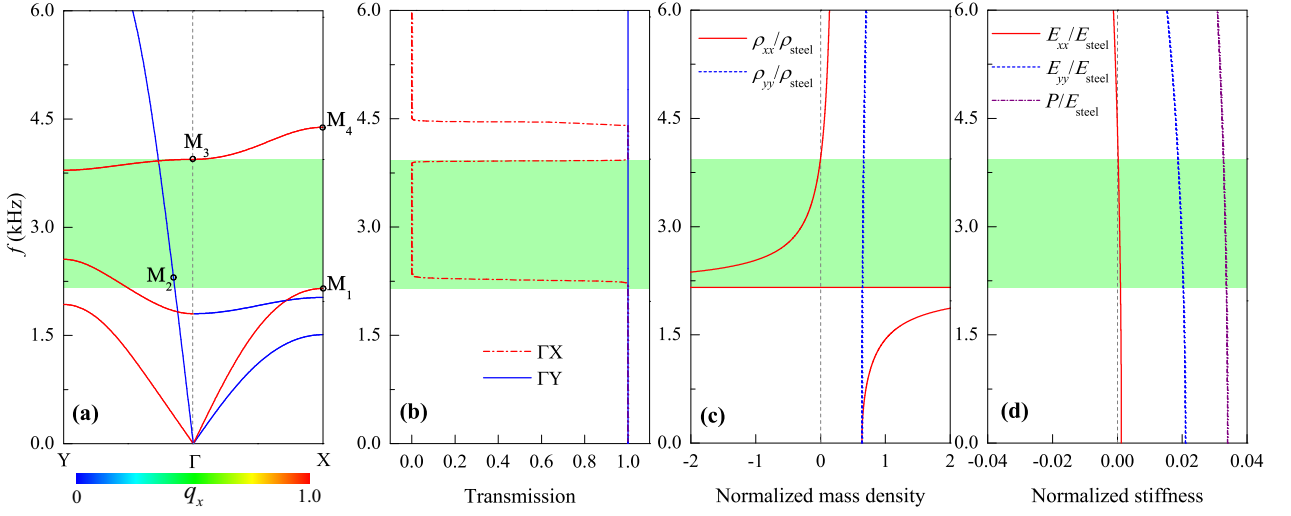

Fig. 4. Characterizations of HEMM S3 in Fig. 2. (a) Band structure along ΓX and ΓY directions for the in-plane waves. (b) Transmission coefficients along two principle directions of a finite HEMM sample for the longitudinal input excitation. (c) Effective mass density along $x$- (red solid line) and $y$- (blue dash line) directions. (d) Effective elastic stiffness ($E_{xx}$: red solid line, $E_{yy}$: blue dash line and $P$: purple dash dot line).

To reveal the physical mechanism of the negative properties, we investigate the representative eigenstates M1-M4 marked in Fig. 4(a), see Fig. 5. The eigenstate $M_1$ in the lower edge of the bandgap has energy mostly concentrated in the six solid blocks. While the eigenstate $M_3$ in the upper edge shows the opposite vibrations. Therefore, the origin of the bandgap in Fig. 4(a) is the result of the enhanced multipolar resonances which produce the negative $\rho_{xx}$ within the range of (2.157 kHz, 3.943 kHz). The eigenstate $M_4$ shows the rotations of the four smaller blocks with the bigger two almost unmoving. This is the typical quadrupolar resonance generating negative $E_{xx}$ above 4.385 kHz as shown in Fig. 4(d). As for the ΓY direction, we also display the eigenstate $M_2$ of the longitudinal wave band within the range of (2.157 kHz, 3.943 kHz). It is shown that the $y$-polarized translation dominates the whole motion of the unit-cell. From these analyses, we can conclude that the optimized HEMM can easily control $x$- and $y$-polarized wave motions independently with the multipolar resonances.

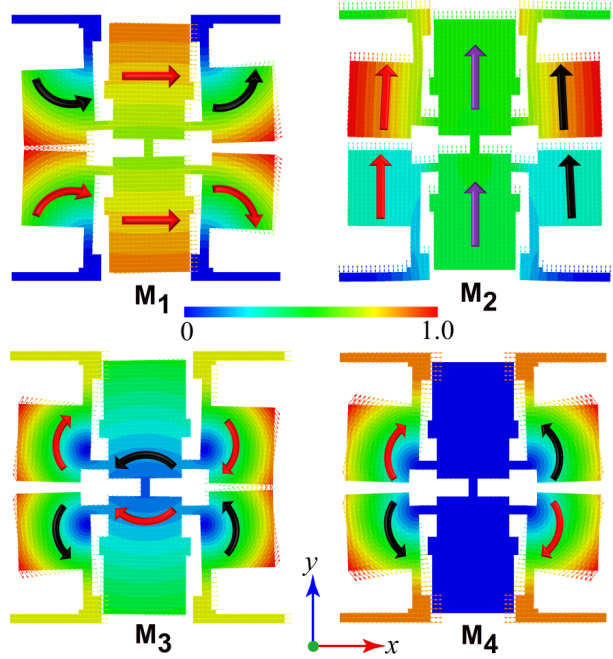

Figure 5. Field distributions of the eigenstates marked in in Fig. 4(a).

*3.3. Subwavelength imagings*

To confirm the hyperbolic dispersion, we illustrate the EFCs of HEMMs S1 and S3 in Figs. 6(a) and 6(b) respectively, leaving details of S2 to the Supplemental Material [36]. Due to the strong anisotropy, the EFCs show the distinctive hyperbolic shape. In this case, the refracted group velocity which is perpendicular to the contours and pointing away from the interface, must be in a negative direction. That is, the negative refraction for the propagating wave mode appears at the interface between the HEMM and background material (stainless steel). It is observed from Figs. 6(a) and 6(b) that the curvature of the curve gets larger with the frequency increasing. Particularly, the bottoms of the hyperboloids clearly show the extremely flat profiles in a broadband frequency range, which can contribute to the energy funneling phenomenon [3, 4] with a large bandwidth. And the flatter curves over the whole wave vector range give rise to the larger group velocities (see Fig. SF3(b) in the Supplemental Material [36]). Furthermore, the three optimized HEMMs (S1-S3) possess the broadband hyperbolic dispersions with bandwidths of 10.651 kHz (S1), 4.095 kHz (S2) and 1.755 kHz (S3). No matter for the subwavelength or deep-subwavelength scales, all these values surpass the reported frequency bandwidths of HEMMs [10, 11]. Certainly, we can freely scale up or down the optimized microstructures for operation at much lower or higher frequencies.

To validate the potential of imaging in optimized HEMMs, we numerically demonstrate typical hyperlensing of the longitudinal waves at various operating frequencies, see Figs. 6(c)-6(f). We work with the optimized HEMMs S1 (Fig.s 6(c) and 6(d)) and S3 (Figs. 6(e) and 6(f)). A slab with 35×8 HEMMs surrounded by the background material is considered. In all cases, a longitudinal-wave point source is assumed located in the position 0.02 m away from the upper side of the HEMM slab. We consider operating frequencies of 13 kHz (Fig. 6(c)), 14 kHz (Fig. 6(d)) for S1 and 2.3 kHz (Fig. 6(e)), 3.1 kHz (Fig. 6(f)) for S3. We observe clearly that the wave propagates through the HEMM and yields an image of the source on the other side. We record the full width at the half maximum (FWHM) of the four images as $0.178\lambda$ (Fig. 6(c)), $0.167\lambda$ (Fig. 6(d)), $0.0156\lambda$ (Fig. 6(e)) and $0.0253\lambda$ (Fig. 6(f)). Moreover, the smaller the operating frequency is, the higher imaging resolution the optimized HEMM can realize. Surprisingly, all these imaging resolutions which are much smaller than the diffraction limit precede the reported values of the HEMMs proposed by Oh et al. [8], Zhu et al. [10] and Lee et al. [11]. The exciting super-high resolution of 0.0156$\lambda$ (~$\lambda$/64) represents the record even in the field of EMMs. We believe that these hyperlensing properties are realized owing to the hyperbolic dispersions with the extremely anisotropic mass densities. In addition, the Fabry-Pérot resonance [35, 36] for the standing wave excitation also can help the hyperlens to improve the imaging transmission at a certain extent.

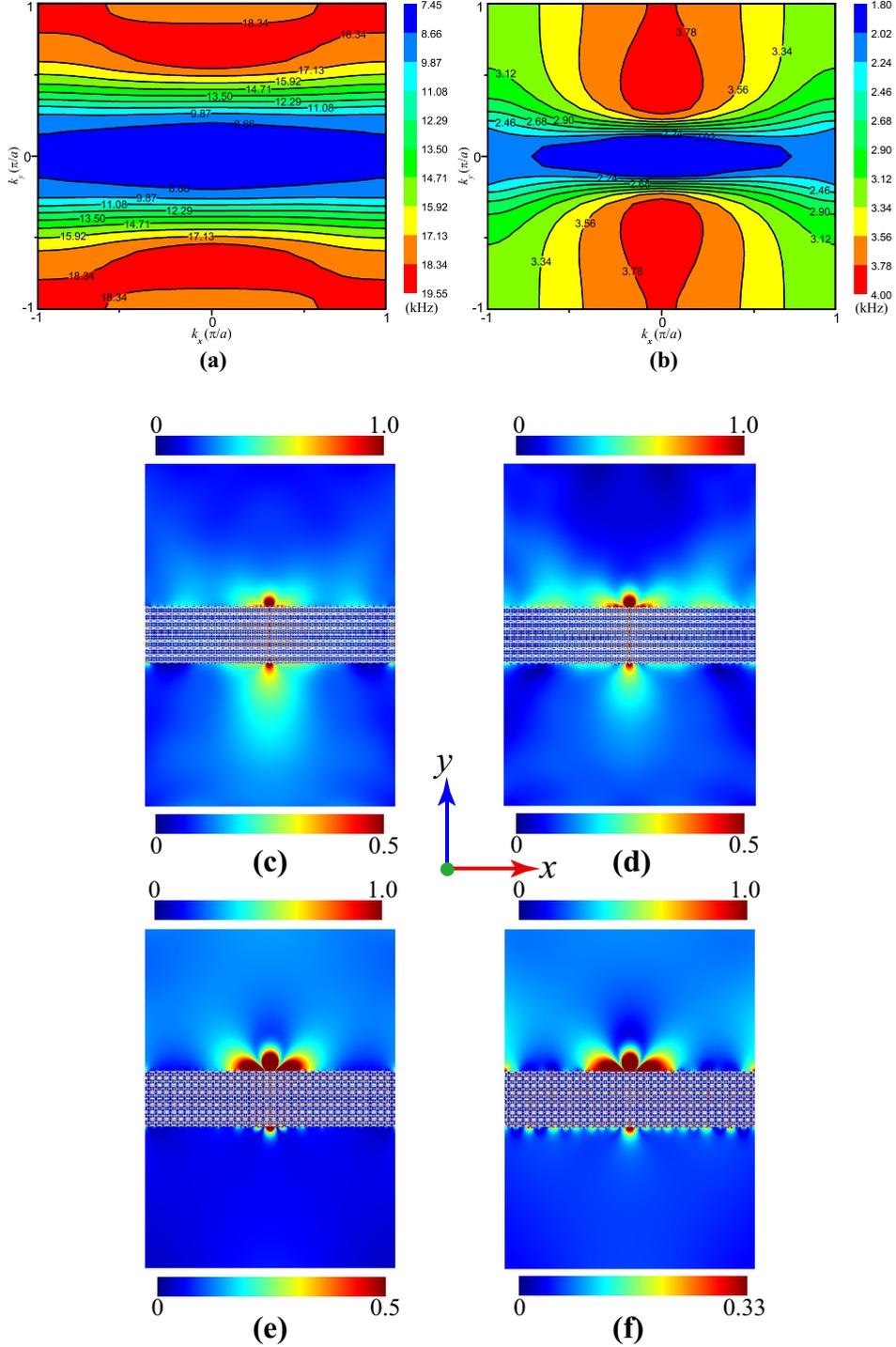

Fig. 6. EFCs and imaging simulations based on HEMMs S1 and S3 in Fig. 2. (a) and (b) EFCs of the third bands for representative optimized microstructures S1 and S3 in (a). (c)-(f) Magnitude filed patterns of longitudinal wave component showing imaging for an 35×8 slab based on optimized structures S1 (c, d) and S3 (e, f) in Fig. 2 at frequencies of (13 kHz, 14 kHz) and (2.3 kHz, 3.1 kHz). The imaging resolutions of simulations in (c), (d), (e) and (f) are FWHM=0.178$\lambda$, 0.167$\lambda$, 0.0156$\lambda$ and 0.0253$\lambda$, respectively. The source is located in the position 0.02 m away from the upper side of HEMM slab.

Since the hyperbolic dispersion is responsible for the above hyperlensing, we present the longitudinal-wave propagating within the optimized HEMMs S1 and S3 at 13 kHz and 2.3 kHz, respectively, to verify the strong anisotropic wave motions, see Fig. 7. We put a point source of the longitudinal wave in the center of a 11×11 HEMM slab. As expected, two images occur at the upper and lower sides of the slabs. However, no visible wave is found at the left and right boundaries. This implies that the longitudinal wave propagates only along $y$-direction, coinciding with the hyperbolic EFCs. It is very interesting to note that the images for these two cases occupy the nearly commensurate regions, suggesting the fact that the optimized HEMMs have the stable focused energy at those frequencies with very flat EFCs. And all widths of the four energy concentration areas are about $1.5a$. We argue that this is due to the similar structure boundaries which can transmit the similar wave motions at the interfaces between the metamaterial and background material. However, from the viewpoint of optimization, it is unlikely to further change the topologies of the boundaries in the optimized metamaterials. Therefore, the similar capacities of wave focusing in Fig. 7 imply that the three HEMMs in Fig. 2 may have the similar ultra-strong imaging abilities for respective wavelength scales. We finally stress that the resolution of $\lambda/64$ shown in Fig. 6(e) possibly represents the hyperlensing performance approaching to the limit under the proposed optimization framework.

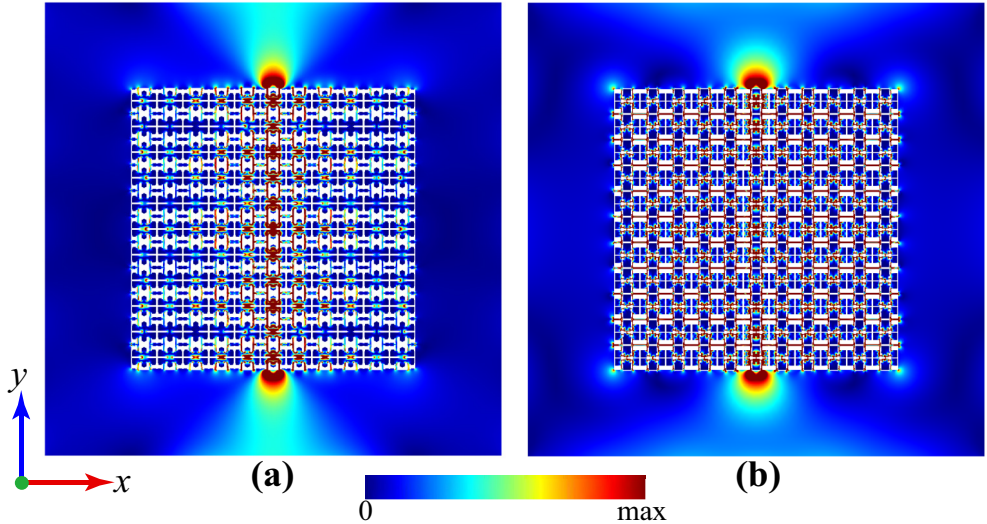

Fig. 7. Wave propagations within HEMMs S1 and S3 in Fig. 2. Magnitude field patterns of the longitudinal wave component for the longitudinal wave propagation in the optimized HEMMS S1 (a) and S3 (b) in Fig. 2 at frequencies of 13 kHz and 2.3 kHz. The resolutions of the images at the upper and lower boundary areas in (a) and (b) are FWHM=$0.169\lambda$ and $0.02\lambda$, respectively. A longitudinal-wave point source is located in the center of the 11×11 HEMM slab.

To reveal the reason of the super resolution, we consider the zero-order transmission coefficient $T$ (the details of $T$ are given in the Supplemental Material [36]) of an elastic plane wave and qualitatively evaluate both propagating and evanescent waves transmitting through a layer (thickness is $8a$) of the optimized HEMM from the free space. Figures 8(a) and 8(b) illustrated the transmission coefficients of the effective anisotropic materials in HEMMs S1 and S3 over certain frequency ranges as the wave component $k_x$ increases. Here, we state that HEMMs S1 and S3 get the negative $\rho_{xx}$ within the ranges of (8.853 kHz, 21.752 kHz) and (2.1576 kHz, 3.9422 kHz), respectively. Meanwhile, both HEMMs have a nearly constant longitudinal wave modulus $P$. From Fig. 8(a), we can clearly observe that the transmission coefficient is improved for a large $k_x$ at the frequency below the negative range. However, owing to the hyperbolic dispersion, the transmission coefficient for the evanescent regions of the $k$-space is relatively large over the whole negative $\rho_{xx}$ range. Obviously, the large transmission coefficient at a high frequency cannot occur for the wide continuous wave vector regions. The global transmission coefficients decrease gradually as the frequency rises. We can notice the analogical performance of the transmission coefficient in Fig. 8(b). At a certain operating frequency below the negative $\rho_{xx}$ range, the transmission coefficient is large only in a narrow wave vector region, especially when the frequency approaches 2.1576 kHz. In particular, the transmission coefficient shows a pronounced increase around 2.5 kHz in a wide $k_x$ range. Therefore, the results in Figs. 8(a) and 8(b) demonstrate that the evanescent waves are enhanced significantly in the optimized HEMMs S1 and S3. Spontaneously, when preforming the focusing, the optimized HEMM with the hyperbolic dispersions will convert the evanescent wave components containing subwavelength information into the propagating components and transfer the energy to the image focal plane. In other words, it is the extreme enhancement of the evanescent waves that yields the super-high imaging resolution of $\lambda/64$.

To explicitly show the effect of the frequency on the transmission, we exhibit in Fig. 8(c) the transmission coefficient in the optimized HEMM S3 at four frequencies (2300 Hz, 2662.5 Hz, 3075 Hz and 3487 Hz). The corresponding effective mass densities in $x$-direction are $\rho_{xx}$=-24526 kg m$^{-3}$, -5303 kg m$^{-3}$, -1948 kg m$^{-3}$ and -698 kg m$^{-3}$, respectively. For comparison, the result for the referential case with the free space at the frequency of 2300 Hz is presented. We define a propagation constant of the fundamental waveguide mode as $k_0=2\pi/\lambda$ [35]. If $k_x \leq k_0$, the transmission coefficient characterize the transmission property for the propagating waves. For $k_x > k_0$, the waves represent the evanescent waves [9]. From the result of the free space, we can easily distinguish evanescent wave transmission with fast attenuation in the free space from the total propagating wave transmission. The other following observations can be made from the results in Fig. 8(c). Because of the imperfect effective impedance matching, the optimized metamaterials in four cases have lower transmission coefficients for the propagating waves than the referential result. However, the strong enhanced transmission of the evanescent waves is excited simultaneously for these four cases. This implies that, contrary to the free space, the optimized HEMM has the competence of generating high imaging resolution. Moreover, as the frequency increases, the transmission of the propagating waves gradually decreases resulting from the worse effective impedance matching. The enhancement extent of the evanescent waves reduces synchronously because the smallest transmission coefficient in the evanescent region decreases obviously. The case at 2300 Hz prominently shows the best and extreme enhancement property over the considered wave vector range. So, the simultaneously best imaging transmission and resolution of the optimized HEMM S3 generally exist at lower frequencies within the hyperbolic frequency range. The other two HEMMs (S1 and S2) also have similar feature. In order to examine the effect of transmission of the propagating and evanescent waves on imaging, we show in Fig. 8(d) the performance of the imaging resolution and maximal intensity on the focal plane based on the constructed HEMM (S3) slab consisting of 35×8 unit-cells at different operating frequencies. Lower resolution (larger FWHM) and lower imaging transmission (smaller maximal intensity) are observed at higher frequencies. We believe that the abovementioned extreme enhancement of the evanescent waves originates physically from the fact that the negative $\rho_{xx}$ is induced by the multipolar resonances which can essentially make $x$- and $y$-polarized vibrations coupled together. Then the evanescent waves can be easily transformed into the propagating wave mode.

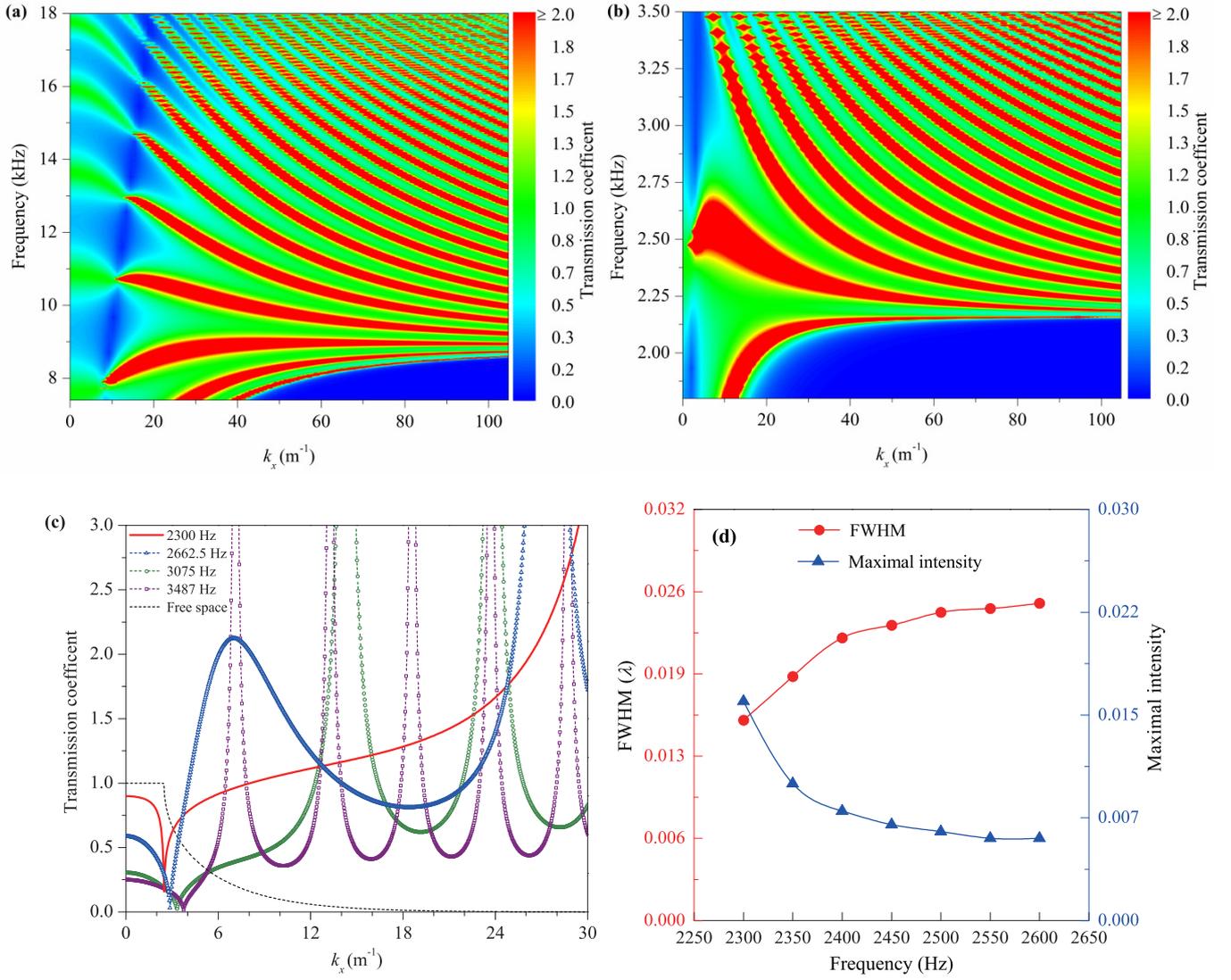

Fig. 8. Enhancement of the evanescent waves and imaging features. Frequency and wave-vector dependence of the transmission through a layer (thickness is 8*a*) of optimized HEMM S1 (a) or S3 (b) in Fig. 2 for both propagating and evanescent waves (note that the transmission coefficient represents its modulus value). (c) Transmissions evaluated for optimized HEMM S3 at several frequencies (2300 Hz: red solid line; 2662.5 Hz: blue dash line with triangles; 3075 Hz: olive dash line with circles; 3487 Hz: purple dash line with squares) and result for the free space at 2300 Hz (black dash line). (d) Performance of the imaging resolution and maximal intensity on the focal plane based on the constructed HEMM slab consisting of 35×8 structures S3 at different operating frequencies.

## 4. Conclusions

Based on the effective mass density, we develop the formulation of the topology optimization of the single-phase metal metamaterial with a broadband hyperbolic dispersion for longitudinal waves. It is shown that the special multipolar resonances guarantee the occurrence of the strong anisotropic effective mass density with a negative value along one direction in the deep-subwavelength frequency region ($\lambda/a \approx 10 \sim 90$). The representative structural topology eventually guides the engineering of HEMMs and even more metamaterial devices with complex functionalities. All imaging simulations of the longitudinal waves through the optimized hyperlens corroborate the ability of transferring subwavelength information, and thereby acquiring the extremely high resolutions beyond the diffraction limit. Moreover, profiting from the extreme enhancement of the evanescent wave transmission, a super-high imaging resolution of about $\lambda/64$ is obtained based on the optimized HEMM at the ultra-low frequency level. Owing to the exploration ability of the topology optimization, the optimized HEMMs in this paper reveal the undiscovered topologies, the record broadband frequency ranges and the record imaging resolutions in the field of EMMs. Overall, the present optimization methodology and results are expected to open the way towards high performance metamaterials in potential applications of medical imaging, sensing and nondestructive testing.

## Acknowledgements

This research is supported by the National Natural Science Foundation of China (Grant No. 11532001), the Chinese Scholarship Council (CSC) and the German Academic Exchange Service through the Sino-German Joint Research Program (PPP) 2014.

## References

[1] Smith, D. R., & Schurig, D. Electromagnetic wave propagation in media with indefinite permittivity and permeability tensors. *Phys. Rev. Lett.* **90,** 077405 (2003).

[2] Poddubny, A., Iorsh, I., Belov, P., & Kivshar, Y. Hyperbolic metamaterials. *Nat. Photonics* **7,** 948-957 (2013).

[3] García-Chocano, V. M., Christensen, J., & Sánchez-Dehesa, J. Negative refraction and energy funneling by hyperbolic materials: An experimental demonstration in acoustics. *Phys. Rev. Lett.* **112,** 144301 (2014).

[4] Christiansen, R. E., & Sigmund, O. Experimental validation of systematically designed acoustic hyperbolic meta material slab exhibiting negative refraction. *Appl. Phys. Lett.* **109,** 101905 (2016).

[5] Liu, Z., Lee, H., Xiong, Y., Sun, C., & Zhang, X. Far-field optical hyperlens magnifying sub-diffraction-limited objects. *Science* **315,** 1686-1686 (2007)..

[6] Lu, D., & Liu, Z. Hyperlenses and metalenses for far-field super-resolution imaging. *Nat. Commun.* **3,** 1205 (2012).

[7] Christensen, J., & de Abajo, F. J. G. Anisotropic metamaterials for full control of acoustic waves. *Phys. Rev. Lett.* **108,** 124301 (2012).

[8] Oh, J. H., Seung, H. M., & Kim, Y. Y. A truly hyperbolic elastic metamaterial lens. *Appl. Phys. Lett.* **104,** 073503 (2014).

[9] Shen, C., Xie, Y., Sui, N., Wang, W., Cummer, S. A., & Jing, Y. Broadband acoustic hyperbolic metamaterial. *Phys. Rev. Lett.* **115,** 254301 (2015).

[10] Zhu, R., Chen, Y. Y., Wang, Y. S., Hu, G. K., & Huang, G. L. A single-phase elastic hyperbolic metamaterial with anisotropic mass density. J. Acoust. Soc. Am. **139,** 3303-3310 (2016).

[11] Lee, H., Oh, J. H., Seung, H. M., Cho, S. H., & Kim, Y. Y. Extreme stiffness hyperbolic elastic metamaterial for total transmission subwavelength imaging. *Sci. Rep.* **6,** 24026 (2016).

[12] Otomori, M., Yamada, T., Izui, K., Nishiwaki, S., & Andkjær, J. Topology optimization of hyperbolic metamaterials for an optical hyperlens. *Struct. Multidisc. Optim.* doi:10.1007/s00158-016-1543-x (2016).

[13] Guo, Y., Cortes, C. L., Molesky, S., & Jacob, Z. Broadband super-Planckian thermal emission from hyperbolic metamaterials. *Appl. Phys. Lett.* **101,** 131106 (2012).

[14] Biehs, S. A., Tschikin, M., & Ben-Abdallah, P. Hyperbolic metamaterials as an analog of a blackbody in the near field. *Phys. Rev. Lett.* **109,** 104301 (2012).

[15] Guo, Y., & Jacob, Z. Thermal hyperbolic metamaterials. *Opt. Express* **21,** 15014-15019 (2013).

[16] Kruk, S. S., Wong, Z. J., Pshenay-Severin, E., O'Brien, K., Neshev, D. N., Kivshar, Y. S., & Zhang, X. Magnetic hyperbolic optical metamaterials. *Nat. Commun.* **7,** 11329 (2016).

[17] Sreekanth, K. V., Alapan, Y., ElKabbash, M., Ilker, E., Hinczewski, M., Gurkan, U. A., De Luca, A., & Strangi, G. Extreme sensitivity biosensing platform based on hyperbolic metamaterials. *Nat. Mater.* **15,** 621-627 (2016).

[18] Poddubny, A. N., Belov, P. A., Ginzburg, P., Zayats, A. V., & Kivshar, Y. S. Microscopic model of Purcell enhancement in hyperbolic metamaterials. *Phys. Rev. B* **86,** 035148 (2012).

[19] Tumkur, T. U., Gu, L., Kitur, J. K., Narimanov, E. E., & Noginov, M. A. Control of absorption with hyperbolic metamaterials. *Appl. Phys. Lett.* **100,** 161103 (2012).

[20] Sreekanth, K. V., ElKabbash, M., Alapan, Y., Rashed, A. R., Gurkan, U. A., & Strangi, G. A multiband perfect absorber based on hyperbolic metamaterials. *Sci. Rep.* **6,** 26272 (2016).

[21] Long, C., Yin, S., Wang, W., Li, W., Zhu, J., & Guan, J. Broadening the absorption bandwidth of metamaterial absorbers by transverse magnetic harmonics of 210 mode. *Sci. Rep.* **6,** 21431 (2016).

[22] Smolyaninova, V. N., Jensen, C., Zimmerman, W., Prestigiacomo, J. C., Osofsky, M. S., Kim, H., Xing, Z., Qazilbash, M. M., & Smolyaninov, I. I. Enhanced superconductivity in aluminum-based hyperbolic metamaterials. arXiv:1605.06817 (2016).

[23] Lu, D., Kan, J. J., Fullerton, E. E., & Liu, Z. Enhancing spontaneous emission rates of molecules using nanopatterned multilayer hyperbolic metamaterials. *Nat. Nanotechnol.* **9,** 48-53 (2014).

[24] Kim, J., Drachev, V. P., Jacob, Z., Naik, G. V., Boltasseva, A., Narimanov, E. E., & Shalaev, V. M. Improving the radiative decay rate for dye molecules with hyperbolic metamaterials. *Opt. Express* **20,** 8100-8116 (2012).

[25] Schulz, K. M., Vu, H., Schwaiger, S., Rottler, A., Korn, T., Sonnenberg, D., Kipp, T., & Mendach, S. Controlling the Spontaneous Emission Rate of Quantum Wells in Rolled-Up Hyperbolic Metamaterials. *Phys. Rev. Lett.* **117,** 085503 (2016).

[26] Caligiuri, V., Dhama, R., Sreekanth, K. V., Strangi, G., & De Luca, A. Dielectric singularity in hyperbolic metamaterials: the inversion point of coexisting anisotropies. *Sci. Rep.* **6,** 20002 (2016).

[27] Mirmoosa, M. S., Kosulnikov, S. Y., & Simovski. C. R. Magnetic hyperbolic metamaterial of high-index nanowires. *Phys. Rev. B* **94,** 075138 (2016).

[28] Oh, J. H., Ahn, Y. K., & Kim, Y. Y. Maximization of operating frequency ranges of hyperbolic elastic metamaterials by topology optimization. *Struct. Multidisc. Optim.* **52,** 1023-1040 (2015).

[29] Kaina, N., Lemoult, F., Fink, M., & Lerosey, G. Negative refractive index and acoustic superlens from multiple scattering in single negative metamaterials. *Nature* **525,** 77-81 (2015).

[30] Lai, Y., Wu, Y., Sheng, P., & Zhang, Z. Q. Hybrid elastic solids. *Nat. Mater.* **10,** 620-624 (2011).

[31] Wu, Y., Lai, Y., & Zhang, Z. Q. Elastic metamaterials with simultaneously negative effective shear modulus and mass density. *Phys. Rev. Lett.* **107,** 105506 (2011).

[32] Zhu, R., Liu, X. N., Hu, G. K., Sun, C. T., & Huang, G. L. Negative refraction of elastic waves at the deep-subwavelength scale in a single-phase metamaterial. *Nat. Commun.* **5,** 5510 (2014).

[33] Liu, F., & Liu, Z. Elastic waves scattering without conversion in metamaterials with simultaneous zero indices for longitudinal and transverse waves. *Phys. Rev. Lett.* **115,** 175502 (2015).

[34] Dong, H. W., Zhao, S. D., Wang, Y. S., & Zhang, C. Topology optimization of anisotropic elastic metamaterial with broadband double-negative index. arXiv:1611.01776 (2016).

[35] Zhu, J., Christensen, J., Jung, J., Martin-Moreno, L., Yin, X., Fok, L., Zhang, X., & Garcia-Vidal, F. J. A holey-structured metamaterial for acoustic deep-subwavelength imaging. *Nat. Phys.* **7,** 52-55 (2011).

[36] See the Supplemental Material, which includes retrieves of effective parameters, genetic algorithm implementation, discussions on objective function, constraints, complementary results of the optimized HEMMs, and transmission of propagating and evanescent waves.

[37] Zhou, S., Li, W., Chen, Y., Sun, G., & Li, Q. Topology optimization for negative permeability metamaterials using level-set algorithm. *Acta Mater.* **59**, 2624-2636 (2011).

[38] Piat, R., Sinchuk, Y., Vasoya, M., & Sigmund, O. Minimal compliance design for metal–ceramic composites with lamellar microstructures. *Acta Mater.* **59**, 4835-4846 (2011).

# Topology optimization of broadband hyperbolic elastic metamaterials with super-resolution imaging

Hao-Wen Dong[a,b], Sheng-Dong Zhao[a,b], Yue-Sheng Wang[a,*], Chuanzeng Zhang[b,*]

[a]*Institute of Engineering Mechanics, Beijing Jiaotong University, Beijing 100044, China*
[b]*Department of Civil Engineering, University of Siegen, D-57068 Siegen, Germany*

**Supplemental Information**

1.  **Retrieves of effective parameters**

To describe the dynamic behaviors of elastic metamaterials (EMMs) [1, 2], it is very useful to adopt the effective medium theory [1]. Since topology optimization involves lots of extremely complex geometries, the numerical-based extraction [1] of the effective material parameters are easier than an analytical method. Under the long wavelength assumption [1, 2], we evaluate the effective material parameters of any microstructure by adding the global displacement fields on the unit-cell boundaries with the displacement phase difference being ignored.

According to the Newton's second law, the mass density tensor can be retrieved from [1, 2]

$$\begin{bmatrix} F_x^* \\ F_y^* \end{bmatrix} = -\omega^2 V \begin{bmatrix} \rho_{xx} & 0 \\ 0 & \rho_{yy} \end{bmatrix} \begin{bmatrix} U_x^0 \\ U_y^0 \end{bmatrix}, \qquad \text{(SQ1)}$$

where $F_x^*$ and $F_y^*$ represent the whole induced forces on the boundaries along $x$- and $y$-directions, respectively; $\omega$ is the angular frequency; $V$ denotes the volume of the effective medium; $\rho_{xx}$ and $\rho_{yy}$ are the effective mass densities in $x$- and $y$-directions, respectively; and the applied displacement fields are represented by $U_x^0$ and $U_y^0$. The off-diagonal elements of the mass density tensor are zero because of the orthogonal symmetry of the microstructural unit-cell.

By applying the corresponding unit eigenstates, the effective stiffness $E_{xx}$ can be determined from the energy equivalence between the unit-cell and the effective medium as expressed by

$$\Sigma(F_x^* U_x^*) = E_{xx} V , \qquad \text{(SQ2)}$$

where $F_x^*$ and $U_x^*$ are the induced nodal force and displacement along $x$-direction at the unit-cell boundaries under the applied time-harmonic displacement field of $u_x = xe^{i\omega t}$ on the left and right boundaries and $u_y=0$ on the upper and lower boundaries. The retrieves of $E_{12}$ and $E_{22}$ can be completed in the similar way. It is noted that we focus on the non-propagating waves along one principle direction for designing hyperbolic dispersion. Therefore, the effective stiffness $E_{xx}$ or $E_{yy}$ should be entirely caused by the induced normal motion along $x$- or $y$-direction. In this case, the applied eigenstates are different from the work published by Liu et al. [1]. However, we can also utilize the longitudinal wave modulus $P=K+\mu$ (where $K$ is the effective bulk modulus and $\mu$ is the effective shear modulus) to characterize the whole effective behaviors concerning the longitudinal wave motions. The detailed process for retrieving $K$ and $\mu$ are referenced to the numerical method proposed by Liu et al. [1].

---
[*] Corresponding author. Tel.: +86 10 51688417; fax: +86 10 51682094. *E-mail address:* yswang@bjtu.edu.cn (Y. S. Wang).
[*] Corresponding author. Tel.: +49 271 7402173; fax: +49 271 7404074. *E-mail address:* c.zhang@uni-siegen.de (Ch. Zhang).

## 2. Genetic algorithm implementation

The improved single-objective genetic algorithm (GA) [3, 4] is adopted to solve the optimization problem (Eqs. (1a)-(1h)). Each binary chromosome involved in GA corresponds to a microstructure formed by a coarse grid with 30×30 pixels (square finite elements regarding the material phase 0 or 1). The search space for optimization has the design variables of $2^{N \times N}$. With the orthotropic symmetry, the total number of possible structures is reduced to $2^{N \times N/4}$. In the GA procedure, a random initial population that contains $N_p$ (=30) individuals (chromosomes) is created. The "abuttal entropy filter" for filling up some isolated voids and removing some isolated elements is applied to improve the structural strength. Then objective function value $SN$ is computed for each individual. The constrained optimization formulation is formed after considering all constraint violation cases. If the $i$th individual is a feasible solution, the final fitness evaluation which is equal to $SN$ is defined as

$$fitness_i = SN_i. \tag{SQ3}$$

Otherwise, if the individual cannot meet some constraints, the fitness is determined by

$$fitness_i = \min(SN_1, SN_2, \cdots SN_{NP}) - \sum_{j=1, 2 \cdots S} |cv_j|, \tag{SQ4}$$

where $S$ is the number of the violated constraints; $cv$ represents the violating extent for a certain constraint. The algorithm gradually employs several genetic operations, including the reproduction that conducts tournament selection with the size of the competition group $N_{ts}$ (=18), crossover with the crossover probability $P_c$ (=0.9) and mutation with the mutate probability $P_m$ (=0.03/0.005 for the coarse/fine grid) to generate the offspring population. The elitism strategy [3, 4] which preserves the best individual in the current generation as an elitism and replaces the worst one in the next generation by the elitism is utilized to accelerate optimization. The optimization process is repeated until a fixed large number (e.g. 1000) of generations is finished. Finally, GA produces an optimized microstructure which can be regarded as a "seed" individual for the new round of optimization in a finer grid with 60×60 pixels for the better description of the structural boundaries. After 1000 iterations, the final optimized microstructure is created.

## 3. Discussion on objective function

Inspired by the mechanism that the dipolar resonances [1, 2, 9] can produce the negative mass density, we should search for the topology of the microstructure to induce a resonance at a certain frequency. The effective value of $\rho_{xx}$ increases to the positive infinity when the operating frequency below the resonant frequency rises. However, the negative $\rho_{xx}$ occurs when the operating frequency above the resonant frequency increases. As a result, the negative $\rho_{xx}$ with a certain frequency width is obtained by the topology optimization scheme [3, 4]. To get the wider negative range at the lower-frequency region, it is necessary to push down the resonant frequency and make the overall positive $\rho_{xx}$ smaller. If we retrieve the effective parameters at the sampling frequencies which are equably distributed in the target operating frequency range ($f_{min}$, $f_{max}$), then the most important driving force to get negative $\rho_{xx}$ in a wide frequency range is to enlarge the gap between the maximal and minimal positive values at these sampling frequencies to excite the resonance. Therefore, we maximize $-\min(\rho_{xx}^+(m))/\max(\rho_{xx}^+(m))$ to achieve this purpose. Once the resonance take places within the target frequency range, the values of $\rho_{xx}$ at several sampling frequencies become negative. In this case, $N$ is introduced in Eq. (1a) to ensure the individual with negative $\rho_{xx}$ to be completely superior to others, thus guiding the evolution to find out more individuals with larger $N$. Through these ways, topology optimization can explore the metamaterial with the broadband negative mass

## 4. Discussions on constraints

In optimization formulation for the negative $\rho_{xx}$, we introduce seven defined constraints (1b)-(1h) to design the broadband hyperbolic dispersion. Here we give the following relative physical reasons for these constraints.

To realize single-negativity for hyperbolic dispersion, we have to guarantee that the eigenstate motions for positive $\rho_{yy}$, $E_{xx}$ and $E_{yy}$ simultaneously when hunting for the negative $\rho_{xx}$. In addition, $E_{yy}$ should be large than $E_{xx}$ to generate the longitudinal band along $y$-direction over a large frequency range in which a bandgap in $x$-direction is opened. These factors offer an explanation to introduction of Eqs. (1b)-(1d) which can make sure the emergence of the longitudinal wave motions along $y$-direction.

When $\rho_{xx}$ becomes negative from positive, the obvious decrease of both $E_{11}$ and $E_{22}$ is usually found, owing to the increase of voids in the metamaterial. However, the multipolar resonances revealed in this paper can effectively induce not only the negative $\rho_{xx}$ but also the negative $E_{xx}$ (although very small). If $E_{11}$ is positive and close to zero, however, it means that the vibrations in $x$-direction can be easily coupled with the excited longitudinal-wave vibrations along $y$-direction. This results in the complex hybrid vibrations containing the longitudinal and transverse wave motions. That is the longitudinal band along $y$-direction may show the quasi-longitudinal wave motions. Therefore, we introduce the constraint of Eq. (1e) to control coupling stiffness $E_{12}$ in order to obtain the pure longitudinal wave motions in $y$-direction.

Normally, we can find the strong "local rotation" which shows the off-diagonal feature of the effective mass density and stiffness tensors, beyond the classic effective linear elastic theory. In particular, the topology optimization usually involves lots of extremely complex microstructures which can bring more serious "local rotation" problems. Accordingly, we have to introduce the constraint of Eq. (1f) to make sure that the induced behavior is completely the translation motion along $y$-direction when retrieving $\rho_{yy}$. Otherwise, the performance of $\rho_{yy}$ cannot be coincided with the dispersion relation.

Extreme anisotropy is the most important feature of hyperbolic metamaterials. To obtain the extremely anisotropic mass densities along the two principle directions, we can make $\rho_{yy}$ keep a nearly constant value when $\rho_{yy}$ takes the variation from the positive infinity to the negative infinity. Because a nearly constant $\rho_{yy}$ indicates the barely dispersive performance along $y$-direction, the extremely different features for the two directions naturally contribute to the hyperbolic dispersion with the maximal anisotropy. So, the constraint of Eq. (1g) is introduced here to achieve this goal. The numerical tests show that $\delta_\rho=1.37$ can effectively balance the requirements of the extreme anisotropy and large feasible solution space.

As for the constraint of Eq. (1h), it is introduced on the fact that we should design the metamaterial providing enough strength in practical manufacturing. Naturally, this constraint can suppress the mesh-dependency problem [3, 5] encountered in topology optimization.

## 5. Complementary results of the optimized HEMMs

All imaging simulations in this paper are executed by commercial software COMSOL Multiphysics. The other physical quantities, including effective parameters, dispersion relations and transmission analysis, are extracted by ABAQUS.

To demonstrate the hyperbolic dispersion for all optimized HEMMs in Fig. 2, we illustrate in Figs. SF1 and SF2 the corresponding dispersion relations, transmission properties of the longitudinal waves and effective constitutive parameters for HEMMs S1 and S2, respectively. It is clearly seen that the different physical quantities match each other very well. The negative $\rho_{xx}$ with the positive $E_{xx}$ can accurately capture the occurrence of the bandgap along the ΓX-direction. The positive $\rho_{yy}$ with the positive $E_{xx}$ also predict the existence of the longitudinal wave mode in the ΓY-direction. The transmission properties along two principal directions also show that the longitudinal waves cannot propagate within the metamaterial in the ΓX-direction but totally propagate along the ΓY-direction. Unlike the results in Figs. 4(a) and SF1(a), the band structure in Fig. SF2(a) shows that only the longitudinal wave mode exists along the ΓY-direction in the ΓX-directional bandgap range. In addition, we illustrate in Fig. SF3(a) the EFCs of the third band for the optimized HEMMs S2 shown in Fig. 2. We can clearly observe the broadband hyperbolic dispersions. Furthermore, the EFCs become very flat at low frequencies as well. It is noted from Fig. SF3(b) that all HEMMs (S1, S2 and S3) possess the positive group velocities along the ΓY-direction within the target frequency ranges. The HEMM S1 even has both positive and negative group velocities at the high-frequency region corresponding to the two eigenstates at the same freqeucnies: one with positive index and the other with double-negative index. The optimized HEMMs with relatively low frequency ranges usually have smaller group velocities. Since we only focus the longitudinal wave propagation, the optimized metamaterials can act as the HEMMs as long as the EFCs for the longitudinal waves have the hyperbolic shapes. So the clear hyperbolic dispersions shown in Figs. 6 and SF3 suggest that our proposed optimization formulation in Eqs. (1a)-(1h) is robust for the longitudinal waves, no matter the transverse wave propagation exists or not.

To check the effect of $\delta_E$ in Eq. (7), we further perform the optimization with $f_{max}$=19.5 kHz and $\delta_E$=0.2, see Fig. SF4. Compared with HEMM S1 ($f_{max}$=19.5 kHz, $\delta_E$=0.1) in Fig. 2, the HEMM in Fig. SF4(a) shows the different geometry features: a big horizontal solid lump connected with the other four small blocks through thick solid connections. Intuitively, the topology will make metamaterial harder to induce the vibrations in $x$-direction when the longitudinal waves propagate in the metamaterial along $y$-direction. Although the resulting EFCs in Fig. SF4(b) also depict the hyperbolic dispersion, we can observe the different characteristic from those of HEMM S1 in Fig. 2, i.e., the hyperbolic dispersions occur at the higher frequencies for the same target range ($f_{min}$, $f_{max}$). Therefore we select a relative small $\delta_E$ (e.g. 0.1 or 0.05) to obtain metamaterials S1, S2 and S3 in Fig. 2 for the lower-frequency ranges.

High transmission is very important in imaging. Generally, two approaches can improve the imaging transmission. One is to construct the metamaterial with a better impedance matching with the background material. The other is to adjust the thickness of the lens to meet the Fabry-Pérot resonant condition [6] for standing wave excitation. To showing the effectiveness of the latter method, we present the imaging simulation for a 15×35 EMM slab based on the optimized HEMM S1 in Fig. 2 at 12.96 kHz. Clearly, Figure SF5 gives the enhanced imaging transmission. More importantly, the obtained imaging resolution (FWHM=0.074$\lambda$) is much smaller than that (FWHM=0.178$\lambda$) in Fig. 6(c). This improvement results from more standing waves excited in the lens. In fact, it is possible to theoretically increase the thickness of the lens in the optimized HEMM S3 in Fig. 2 for a higher resolution larger than 0.0156$\lambda$.

## 6. Transmission of propagating and evanescent waves

All optimized metamaterials showing outstanding deep-subwavelength imaging benefits from the transimission enhancements of the evanescent waves which can convert the subwavelength information into the propagating components. In order to characterize the transmission [7, 8], we perform the numerical evaluation on the propagating and evanescent waves transmitting through a layer of the anisotropic effective medium from the free space. In this paper, we only focus the hyperbolic dispersion of the longitudinal waves by neglecting the transverse waves. Because of the anisotropic mass density and anisotropic stiffness, it is impossible to drive the hyperbolic form from the elastic wave equation of the super anisotropic effective elastic medium. As a remedy [9], due to the nearly constant value, the longitudinal wave modulus $P$ can be utilized to describe the whole longitudinal wave motion. In this case, the anisotropic mass density dominates the formation of the hyperbolic dispersion for elastic waves. Referring to the study reported by Zhu et al. [9], the dispersion relation for the longitudinal wave propagating within EMMs with the anisotropic mass density can be expressed by

$$\frac{k_x^2}{\rho_{xx}} + \frac{k_y^2}{\rho_{yy}} = \frac{\omega^2}{P} \tag{SQ5}$$

where $k_x$ and $k_y$ are wave vectors in the effective anisotropic elastic medium along $x$- and $y$-directions, respectively. The transmission ($T$) of the propagating and evanescent waves is defined as [7, 8]

$$T = \frac{4Z_y Z_{0y} e^{ik_y L}}{(Z_y + Z_{0y})^2 - (Z_y - Z_{0y})^2 e^{2ik_y L}} \tag{SQ6}$$

where $Z_y = \omega \rho_{yy}/k_y$ and $Z_{0y} = \omega \rho_0/k_{0y}$ are the wave impedances; $\rho_0$ is the mass density of the background medium; $L$ is the thickness of the layer in the optimized metamaterial; $k_{0y} = \sqrt{k_0^2 - k_x^2}$ is the wave vector in the free space; and $k_0 = 2\pi/\lambda$ denotes the propagating constant of the fundamental waveguide mode [6]. For a microstructure generated from the GA, substituting the effective mass densities ($\rho_{xx}$ and $\rho_{yy}$) and effective longitudinal wave modulus ($P$) into Eq. (SQ6) returns the defined transmission $T$ at a certain frequency.

**References**

[39] Liu, X. N., Hu, G. K., Sun, C. T., & Huang, G. L. Wave propagation characterization and design of two-dimensional elastic chiral metacomposite. *J. Sound Vib.* **330,** 2536-2553 (2011).

[40] Liu, X. N., Hu, G. K., Huang, G. L., & Sun, C. T. An elastic metamaterial with simultaneously negative mass density and bulk modulus. *Appl. Phys. Lett.* **98,** 251907 (2011).

[41] Dong, H. W., Su, X. X., & Wang, Y. S. Multi-objective optimization of two-dimensional porous phononic crystals. *J. Phys. D: Appl Phys.* **47,** 155301 (2014).

[42] Dong, H. W., Su, X. X., Wang, Y. S., & Zhang C. Topology optimization of two-dimensioanl phononic crystals based on the finite element method and genetic algorithm. *Struct. Multidisc. Optim.* **50**, 593-604 (2014).

[43] Sigmund, O., & Petersson, J. Numerical instabilities in topology optimization: a survey on procedures dealing with checkerboards, mesh-dependencies and local minima. *Struct. Multidisc. Optim.* **16,** 68-75 (1998).

[44] Zhu, J., Christensen, J., Jung, J., Martin-Moreno, L., Yin, X., Fok, L., Zhang, X., & Garcia-Vidal, F. J. A holey-structured metamaterial for acoustic deep-subwavelength imaging. *Nat. Phys.* **7,** 52-55 (2011).

[45] Zhou, X., & Hu, G. Superlensing effect of an anisotropic metamaterial slab with near-zero dynamic mass. *Appl. Phys. Lett.* **98,** 263510 (2011).

[46] Shen, C., Xie, Y., Sui, N., Wang, W., Cummer, S. A., & Jing, Y. Broadband acoustic hyperbolic metamaterial. *Phys. Rev. Lett.* **115,** 254301 (2015).

[47] Zhu, R., Chen, Y. Y., Wang, Y. S., Hu, G. K., & Huang, G. L. A single-phase elastic hyperbolic metamaterial with anisotropic mass density. *J. Acoust. Soc. Am.* **139,** 3303-3310 (2016).

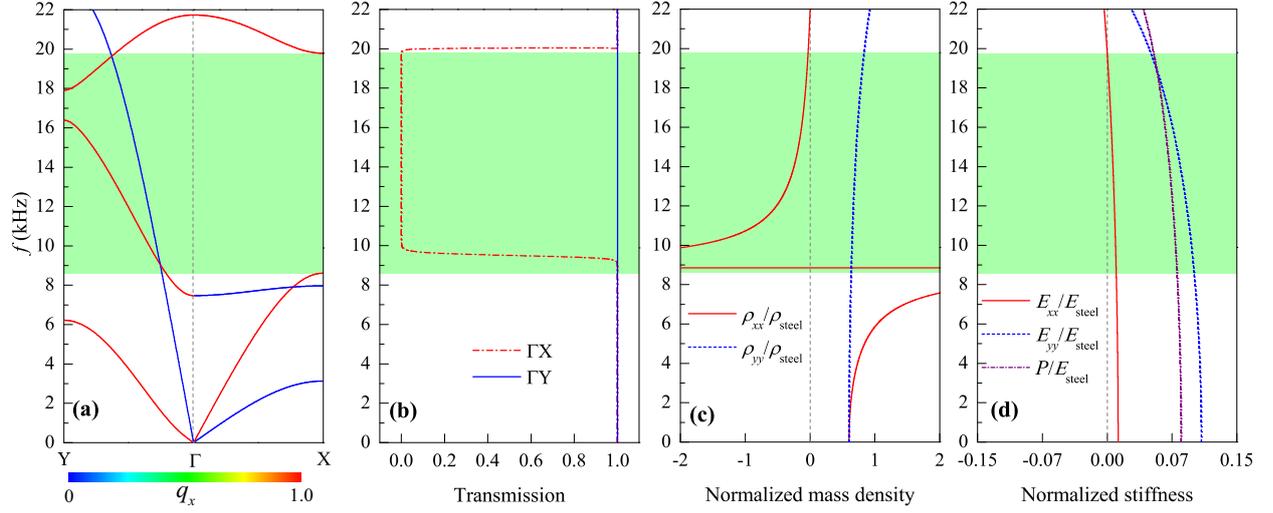

**Fig. SF1.** Characterizations of HEMM S1 in Fig. 2. (a) Band structure along ΓX- and ΓY-directions for the in-plane waves. (b) Transmission coefficients along two principle directions of a finite HEMM sample for the longitudinal input excitation. (c) Effective mass density along $x$- (red solid line) and $y$- (blue dash line) directions. (d) Effective elastic stiffness ($E_{xx}$: red solid line, $E_{yy}$: blue dash line and $P$: purple dash dot line).

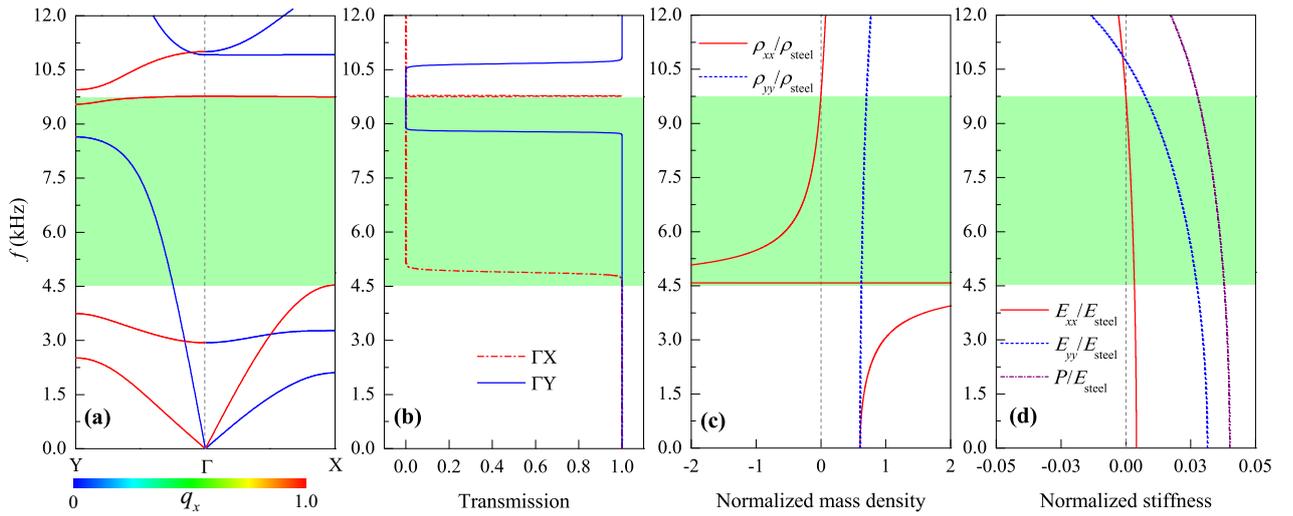

**Fig. SF2.** Characterizations of HEMM S2 in Fig. 2. (a) Band structure along ΓX- and ΓY-directions for the in-plane waves. (b) Transmission coefficients along two principle directions of a finite HEMM sample for the longitudinal input excitation. (c) Effective mass density along $x$- (red solid line) and $y$- (blue dash line) directions. (d) Effective elastic stiffness ($E_{xx}$: red solid line, $E_{yy}$: blue dash line and $P$: purple dash dot line).

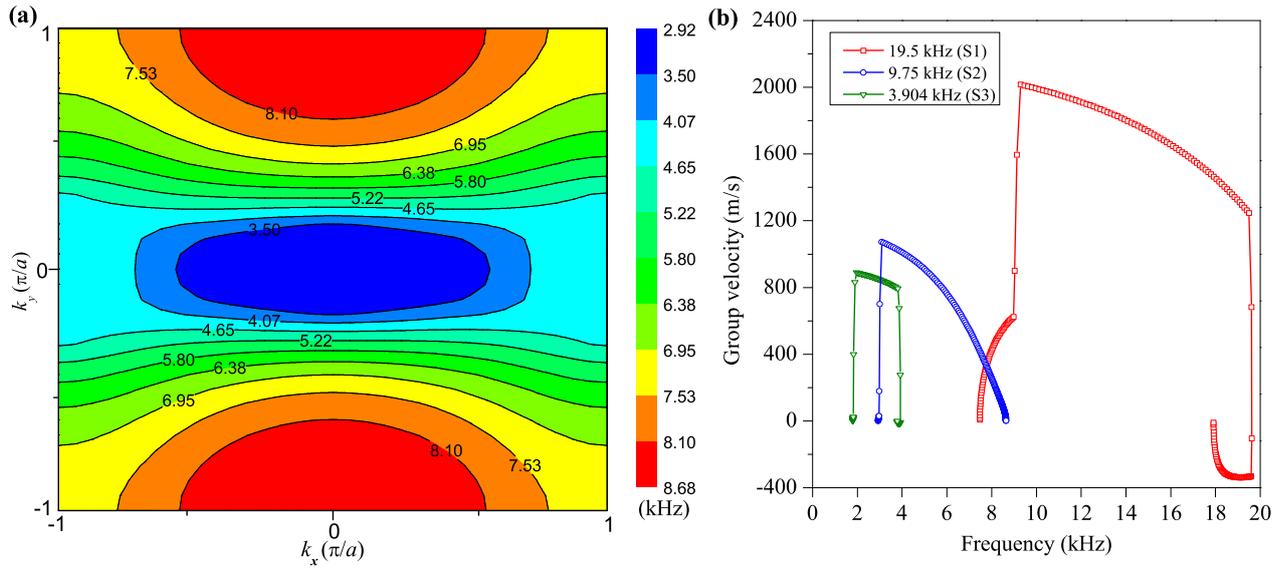

**Fig. SF3.** EFCs of the third band for the representative optimized HEMMs S2 (**a**) in Fig. 2. (**b**) The group velocities of the third band along the ΓY direction for three HEMMs S1 ($f_{max}=f_1=19.5$ kHz), S2 ($f_{max}=f_2=9.75$ kHz) and S4 ($f_{max}=f_3=3.904$ kHz).

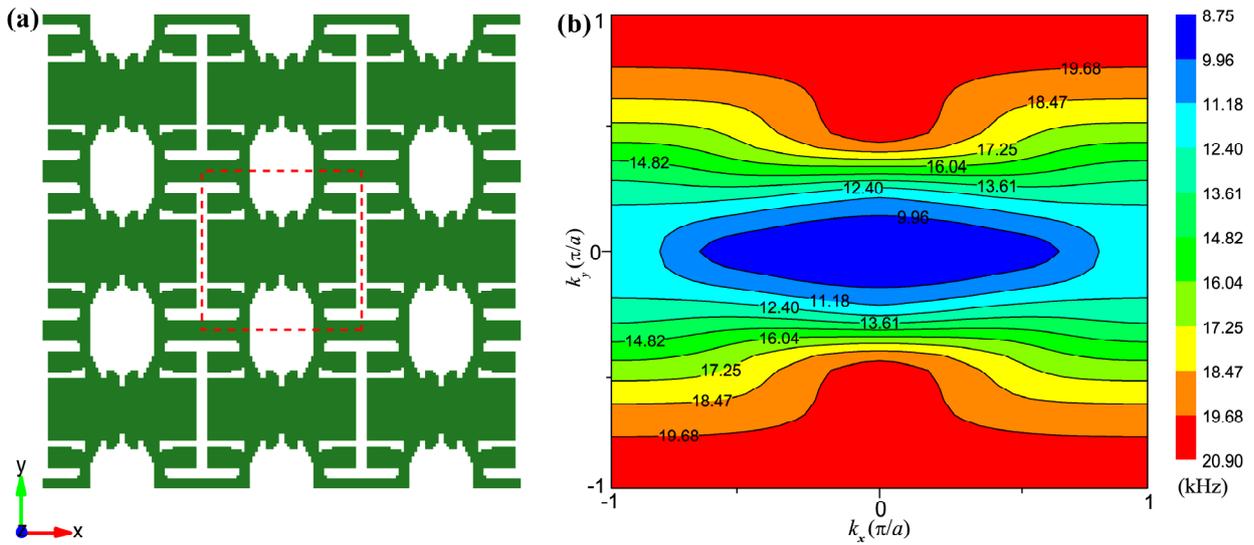

**Fig. SF4.** Optimized HEMM with unit-cell enclosed by the dashed lines ($f_{max}=19.5$, $\delta_E=0.2$) (a) and its EFCs of the third band (b).

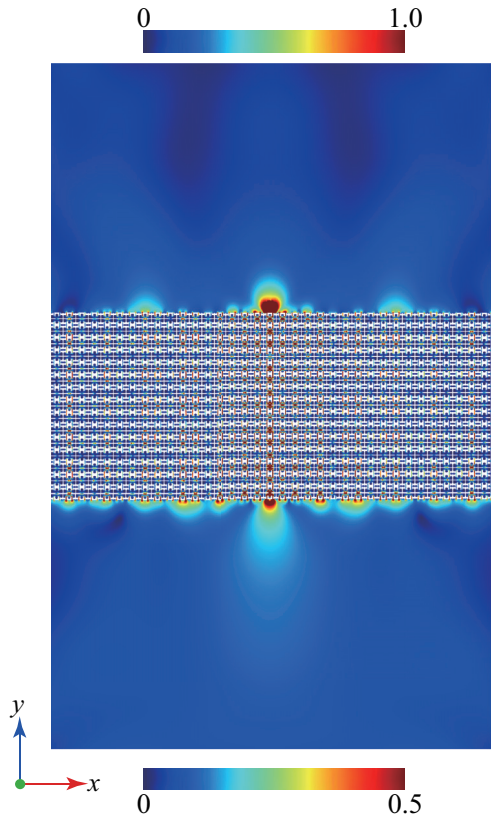

**Fig. SF5.** Magnitude filed patterns of longitudinal wave component showing imaging for an 35×15 EMM slab based on optimized HEMM S1 in Fig. 2 at 12.96 kHz (the Fabry-Pérot resonant condition for standing wave excitation is satisfied). The imaging resolutions is FWHM=0.074$\lambda$.